    \documentclass[12pt,psf,epsf]{article}
\usepackage[centertags]{amsmath}
\usepackage{amssymb}
\usepackage{graphicx}
\usepackage{epsfig}
\usepackage{ulem}
\usepackage[english]{babel}
\usepackage{array}
\usepackage{amsthm}
\usepackage{latexsym}
\usepackage[mathcal]{euscript}
\pdfoutput=1
\usepackage{epsfig}

 \usepackage{jheppub}
 \usepackage{hyperref}

\usepackage{subfigure}
\usepackage{psfrag}

\usepackage{epsfig}
\usepackage[latin1]{inputenc}
\usepackage{float}
\usepackage{graphicx}
\usepackage{cancel}
\usepackage{mathrsfs}
\usepackage{amssymb}
\usepackage{amsfonts}
\usepackage{amsmath}
\usepackage{slashed}
\usepackage{graphicx}
\usepackage{bm}
\usepackage{color}
\newcommand{\be}{\begin{equation}}
\newcommand{\ee}{\end{equation}}
\newcommand{\bea}{\begin{eqnarray}}
\newcommand{\eea}{\end{eqnarray}}
\newcommand{\bwt}{\begin{widetext}}
\newcommand{\ewt}{\end{widetext}}

\newcommand{\nn}{\nonumber}

\newcommand{\bi}{\begin{itemize}}
\newcommand{\ei}{\end{itemize}}

\usepackage{setspace}

\definecolor{darkraspberry}{rgb}{0.53, 0.15, 0.34}

\definecolor{red}{rgb}{1, 0, 0}

\definecolor{darkblue}{rgb}{0., 0, 1}

\definecolor{dgreen}{rgb}{0.,0.6,0.}

\begin{document}

\title {Thermal density matrix breaks down the Page curve}

\author{Dmitry S. Ageev,}

\author{Irina Ya. Aref'eva}
\affiliation{Steklov Mathematical Institute, Russian Academy of Sciences, Gubkin str. 8, 119991
Moscow, Russia}

\emailAdd{ageev@mi-ras.ru}
\emailAdd{arefeva@mi-ras.ru}

\abstract{In this paper, we study entanglement islands and the Page curve in the eternal four-dimensional Schwarzschild black hole surrounded by finite temperature conformal matter. By finite temperature conformal matter we mean the matter described by the thermal density matrix, rather than the usually considered matter above the Fock vacuum. We take the matter and the black hole at different temperatures and calculate the entanglement entropy for such a setup using the s-wave approximation. As a result, we obtain that at late times the island prescription leads to the exponential growth of the entanglement entropy of conformal matter in thermal vacuum.\\

}

\maketitle

\section*{Introduction}
\addcontentsline{toc}{section}{Introduction}
The behavior of quantum matter in curved spacetime is naturally defined by the metric of the underlying manifold. In the non-dynamical cFase of geometry, one can proceed with a standard way to calculate the observable \cite{Birrell:1982ix}. However, for dynamical gravity in some cases one cannot separate the calculations related to the geometry and the matter even when the back-reaction is negligible.  The well-known quantum information measure such as entanglement entropy is an explicit example of such a calculation. This is of particular interest in view of importance of the entanglement entropy for the  study of gravitational entropic objects like black holes  \cite{Haw1,Haw2,Page:1993wv,Page:2013dx}. The fact that one should take care of the gravity part in the entanglement calculation  led to a series of breakthroughs explaining the resolution of the information paradox, namely, to the correct behavior of the entanglement entropy through the emergence of special regions called ``the entanglement islands'' \cite{Penington:2019npb}-\cite{AV}. The emergence of the islands is shown to be driven by the {\it replica wormhole} mechanism which takes into account the presence of dynamical gravity \cite{Penington:2019kki,Almheiri:2019qdq}. The entanglement islands restore the unitary behavior of the entanglement entropy  the time-dependence of which should be described by the so-called Page curve.
 During the time evolution, initially increasing entanglement must at least stop doing it or begin to decrease at some characteristic time $\tau_{P}$ known as the Page time $\tau_{P}$. A detailed description of this setup is possible in two-dimensional gravity \cite{Penington:2019kki,Almheiri:2019qdq} and BCFT/moving mirror models \cite{Rozali:2019day}-\cite{Hartman:2013qma}, however, for the four-dimensional setting one has to use different approximations and assumptions to perform the calculations \cite{Dong:2020uxp}. In \cite{Dong:2020uxp}, the Page curve for a particular setup in eternal four-dimensional Schwarzschild black hole has been considered (see \cite{Krishnan:2020}-\cite{Yu:2021cgi} for different extensions and other studies of the islands in higher-dimensional flat space black holes). An important part of the calculations (see ~\cite{HIM}) is the splitting of the eternal black hole metric into the conformally flat part and the complement. After that, the entanglement of conformal matter is assumed to be given by the two-dimensional CFT answer (for CFT defined on the conformal part of the black hole) or, in other words, by taking the s-wave approximation. In \cite{HIM}, it was shown how the island prescription leads to  the entanglement saturation to  a constant value after some linear growth in this setup. This model being simple, straightforward, and easy to modify is an excellent testing ground for the island prescription. Recently, in \cite{AV}, it was argued how the dynamical setup involving the black hole adiabatic evaporation (i.e. a black hole with changing mass) could spoil the unitarity.\\ 

In this paper, we suggest another setup which leads to a ridiculous form of the Page curve even for the eternal black hole. Roughly speaking, the island formula strongly depends on the matter for which the entanglement is calculated. Typically, one considers the conformal matter placed in the curved space in its ground state. Thermality is the key feature of a black hole, so the matter ground state in such a background also leads to a thermal-like behavior. However, in principle, one can take the thermal averaging in such a theory with respect  to another  density matrix with the different temperature. For example, in \cite{Susskind:1994sm}, two-dimensional models have been studied with the conclusion that while the matter fields could be averaged using thermal density matrix in curved space.  A similar gravitational calculation (in the path-integral formalism) shows some additional divergences, thus, picking out the special temperature value equal to the Hawking one. 
We interpret this setup as the ``thermal matter'' or ``thermal gas'' surrounding the black hole. What we find is that even for small temperatures of the gas one can observe the explosive growth of the entanglement entropy.

The paper is organized as follows. In section \ref{sec:sec2} we introduce our setup and construction of islands in Schwarzschild black hole. In section \ref{sec:sec3} we calculate island dynamics and its effect on thermal gas entanglement.

\section{Islands in Schwarzschild black hole}\label{sec:sec2}
\subsection{Setup with vacuum CFT}
In this paper, we consider Einstein gravity  interacting with some matter with the action $I_{\text {matter}}$. The total action $I$ has the form
\be \label{eq:act}
\begin{aligned}
I &=I_{\text {gravity }}+I_{\text {matter }}, \\
I_{\text {gravity }} &=\frac{1}{16 \pi G_{\mathrm{N}}} \int_{\mathcal{M}} d^{4} x \sqrt{-g} R+\frac{1}{8 \pi G_{\mathrm{N}}} \int_{\partial \mathcal{M}} d^{3} x \sqrt{-h} K,
\end{aligned}
\ee 
where we included the boundary term and denoted by $G_{\mathrm{N}}$ the Newton's gravitational constant. We neglect the backreaction of the matter fields $I_{\text{matter}}$ living on  the  Schwarzschild black hole background  with the metric 
$$
d s^{2}=-\frac{r-r_{\mathrm{h}}}{r} d t^{2}+\frac{r}{r-r_{\mathrm{h}}} d r^{2}+r^{2} d \Omega^{2}.
$$
Here, the horizon is located at $r_{\mathrm{h}}$ and the black hole temperature is given by
$$
T_{\mathrm{H}}=\frac{1}{\beta}=\frac{1}{4 \pi r_{\mathrm{h}}} .
$$
In the Kruskal coordinates, the eternal black hole metric takes the form
\be \label{eq:kruskal}
d s^{2}=-\frac{d U d V}{W^{2}}+r^{2} d \Omega^{2},\,\,\,\,
W=\sqrt{\frac{r}{4 r_{\mathrm{h}}^{3}}} e^{\frac{r-r_{\mathrm{h}}}{2 r_{\mathrm{h}}}},
\ee 
where the coordinates $U$, $V$ and $r_*$ are defined as 
\begin{gather}\label{eq:coord}
r_{*}=r-r_{\mathrm{h}}+r_{\mathrm{h}} \log \frac{r-r_{\mathrm{h}}}{r_{\mathrm{h}}}, \\
U \equiv-e^{-\frac{t-r_{*}}{2 r_{\mathrm{h}}}}=-\sqrt{\frac{r-r_{\mathrm{h}}}{r_{\mathrm{h}}}} e^{-\frac{t-\left(r-r_{\mathrm{h}}\right)}{2 r_{\mathrm{h}}}}, \quad V \equiv e^{\frac{t+r_{*}}{2 r_{\mathrm{h}}}}=\sqrt{\frac{r-r_{\mathrm{h}}}{r_{\mathrm{h}}}} e^{\frac{t+\left(r-r_{\mathrm{h}}\right)}{2 r_{\mathrm{h}}}} .
\end{gather}
Now let us briefly describe the setup used in \cite{HIM} to demonstrate the island influence on the entanglement evolution in a four-dimensional Schwarzschild black hole. We consider two large regions located in the left and right parts of the eternal black hole described by the metric \eqref{eq:kruskal} in coordinates \eqref{eq:coord}. The boundaries of these regions denoted by $R_-$ and $R_+$ are placed at the points $b_-$ and $b_+$ in the left and right wedges respectively. As we mentioned before, an important part of the setting which we consider is the special region called the entanglement island $I$ bounded by the points $a_-$ and $a_+$. More precisely, we choose the parametrization of $a_\pm,b_\pm$ by some points $r=a$ and $r=b$ as follows 
\begin{gather} \label{eq:points}
 a_+=\left(t_{a}, a\right),\,\,\,\, a_{-}=\left(-t_{a}+i \frac{\beta}{2}, a\right),\\
 b_+=\left(t_{b}, b\right),\,\,\,\,b_-=\left(-t_{b}+i \beta / 2, b\right).
 \end{gather}
The geodesic distance between two points $x_1$ and $x_2$ in the geometry defined by \eqref{eq:kruskal} is given by
\be \label{eq:d}
d\left(x_{1}, x_{2}\right)=\sqrt{\frac{\left(U\left(x_{2}\right)-U\left(x_{1}\right)\right)\left(V\left(x_{1}\right)-V\left(x_{2}\right)\right)}{W\left(x_{1}\right) W\left(x_{2}\right)}}.
\ee
In \cite{HIM}, it was shown that the evolution of the entanglement entropy in this setting is given by the competition of two contributions. The first one which typically contributes at early times corresponds to the entropy of
the matter fields 
\be S_1=S_{1}\left(R_{+} \cup R_{-}\right).
\ee
The second one denoted by $S_2$  contributes, being minimal, at late times and its presence is due to the appearance of the recently discovered entanglement island phenomenon. The island formula describing this phenomenon states that a so-called generalized entropy $S_2$ has the form
\be \label{eq:S2gen}
S_2=\min \left\{\operatorname{ext}\left[\frac{\operatorname{Area}(\partial I)}{4 G_{\mathrm{N}}}+S_{\text {matter }}(R \cup I)\right]\right\},
\ee 
where the extremization over all possible islands and taking the minimum value from the considered set is assumed.

Following \cite{HIM}, $S_1$  can be approximated as the entanglement entropy of matter fields in the region constrained by $b_-$ and $b_+$ 
\be \label{eq:S1}
S_1=S(b_-,b_+),
\ee 
 where we denote by $S(x_1,x_2)$ the entanglement entropy of the matter fields located  in the region $(x_1,x_2)$.  
 We assume that in terms of $S(x_1,x_2)$ the generalized entropy has the factorized form
 \be \label{eq:S2}
 S_2=\frac{2 \pi a^2}{G_N}+S(a_+,a_-)+S(b_+,b_-)+S(a_+,b_+)+S(a_-,b_-)-S(a_+,b_-)-S(a_-,b_+)
 \ee 
where the first term  comes from the area contribution $\operatorname{Area}(\partial I)/4 G_{\mathrm{N}}$ in  \eqref{eq:S2gen} for the metric given by \eqref{eq:kruskal}.\\

An important part of the island calculation is the knowledge of the $S(x_1,x_2)$ explicit form. The calculation of entanglement entropy for a general quantum theory in a finite region  is a challenging problem. Explicit results are available in their essence only for two-dimensional CFT where the infinite conformal symmetry simplifies the calculation. In our setup, we have a higher-dimensional curved spacetime with a non-trivial structure which makes the problem nearly unsolvable at first sight (at least analytically) even for a massless fields. Moreover, as the gravity is assumed to be dynamical (i.e. the black hole is the solution of the Einstein gravity) one should be careful when calculating the quantum entanglement (see for example \cite{Dong:2020uxp} for the consideration of the entanglement effective theory).\\

The last important piece of the general setting we use is a natural and reasonable assumption which has been suggested in \cite{HIM} --- the so-called ``s-wave approximation'' for the entanglement entropy $S(x_1,x_2)$. In this approximation, we neglect the spherical part of the metric \eqref{eq:kruskal} and consider only the conformally flat part. Then one can use 2d-CFT intuition where the entanglement entropy of the interval with the endpoints $x_1$, $x_2$ is just a function of the distance between the points: $x_2-x_1$. In the ground state on the conformally flat background, $S(x_1,x_2)$ is approximated by the replacement
\be \label{eq:Sintro}
S_{CFT}(x_1,x_2)=\frac{c}{3}\log \frac{(x_2-x_1)}{\varepsilon} \rightarrow S(x_1,x_2)=\frac{c}{3}\log \frac{d(x_1,x_2)}{\varepsilon},
\ee 
where $S_{CFT}$ is the entanglement entropy of the interval in the 2d-CFT ground state, and $d (x_1, x_2)$  is given by \eqref{eq:d}.
Within the usage of the formula \eqref{eq:Sintro}, the initial growth and subsequent saturation at the Page time has been derived in a four-dimensional flat black hole setting in \cite{HIM}.

\subsection{Setup with thermal gas}

The thermality property is intrinsic to the black hole background. We consider the generalization of \eqref{eq:Sintro} to the case when the black hole interacts with the conformal matter at a finite temperature. Before introducing the specific formula which we will use, several comments are in order. Typically, one performs the calculation with a matter field in the ground state with respect to the black hole geometry and then after some manipulations, one can show that \ observers at spatial infinity will detect thermal radiation. Using the gravitational path-integral in calculations leads to a particular value of the temperature equal to the Hawking one. In principle, one can follow another way and calculate the expectation value with respect to the thermal density matrix (for some operator  ${\cal O}$) with respect to the matter Hamiltonian $H$ in a standard way as $\text{Tr}\langle  {\cal O} e^{-\beta H} \rangle$. Here, one can observe the presence of freedom in the choice of the matter temperature $T=1/\beta$. For example, in \cite{Susskind:1994sm}, two-dimensional Rindler space with thermal matter has been considered in this way. It was shown that while calculating simple quantities (like the free energy) for the single-mode states with fixed $\beta$ different from the Rindler temperature, nothing special happens. Proceeding with a similar procedure for gravity in the path-integral formalism leads to some drawbacks like the presence of numerous divergences due to the conical singularity. When Rindler and the gas temperature coincide we get a single divergence which is to be eliminated by the gravitational constant renormalization (which is in line with the previously noticed similar elimination of the divergence in the entanglement \cite{HIM}).  In general, the presence of the divergence is natural since we consider the entanglement entropy in quantum field theory. In two-dimensional JT gravity model from \cite{Almheiri:2019yqk,Almheiri:2019qdq}, the temperature of the CFT bath is in equilibrium with the black hole system and fixed by some equilibrium condition. In general, one considers our setup as a slowly evolving gas surrounding the black hole, so extremely long times are not considered here. Taking all this into account for the matter entanglement entropy, one can convince themselves of the possibility of the existence of different temperatures of the gas and the black hole during a reasonable amount of time. In two-dimensional  CFT in flat space, the von Neumann entropy of a single interval is well known and given by 
\be \label{eq:ST}
S_{CFT}(x_1,x_2)=\frac{c}{3} \log \left(\frac{\sinh \left(\pi T\left(x_{2}-x_{1}\right)\right)}{\pi T \varepsilon}\right).
\ee
Similarly to \eqref{eq:Sintro}, this leads to the generalization of the entanglement by replacement $|x_2-x_1|\rightarrow d(x_1,x_2)$. We use this formula as a conjectural statement naturally generalizing the zero-temperature formula. One of the ways to consider the finite temperature entanglement in curved spacetime is to construct a  topological black hole over the metric of interest. The explicit form of this dependence is
\be \label{eq:STd}
S(x_1,x_2)=\frac{c}{3} \log \left(\frac{\sinh \left(\pi T\cdot d\left(x_{1},x_{2}\right)\right)}{\pi T \varepsilon}\right)
\ee
where $T$ is the temperature of matter fields. In general, we assume that $T_H \neq T$ and associate $T_H$ with the temperature of the black hole background calculated from the gravitational path integral.  Let us calculate the explicit form of the entanglement entropy evolution and explore the effect caused by the finite temperature matter.

As a last remark, we would like to notice that, as we mentioned before, in the two-dimensional dilaton gravity calculation \cite{Almheiri:2019qdq}, the formula is slightly different and obtained by a subsequent mapping of the hyperbolic disk geometry metric (equipping with a black hole) and the Weyl rescaling. It is not clear whether such an approach is applicable here, especially due to the presence of additional ``spherical'' degrees of freedom which we neglect. The difference comes because the expression under the hyperbolic sine (in the numerator) depends on $\left(U\left(x_{2}\right)-U\left(x_{1}\right)\right)\left(V\left(x_{1}\right)-V\left(x_{2}\right)\right)$, while the conformal factors $W$ are still present in the denominator. The main effects described in the further sections qualitatively depend only on the expression in $\sinh$.

\section{Entanglement islands  contributions in heated environment}
\label{sec:sec3}
\subsection{Late time behaviour}
Before considering the initial growth  defined by a rather simple expression \eqref{eq:S1}, let us calculate the entanglement island contribution dominating at late times where we expect to have a more complicated dynamics. 
The island contribution has the form \eqref{eq:S2} where the points $a_\pm$ and $b_\pm$ are defined by \eqref{eq:points}, the distance is given by \eqref{eq:d} and the entanglement entropy is \eqref{eq:STd}. The total expression for the generalized entropy calculated by these formulae has quite a cumbersome form, however, introducing the shift $\delta=t_a-t_b$ and the time $t_b=t$, the asymptotics at large times $t\rightarrow \infty$ can be evaluated explicitly as the sum of four ${\cal S}_i$-terms
\be \label{eq:S22}
S_2=\frac{2 \pi a^2}{G_N}+{\cal S}_1+{\cal S}_2+{\cal S}_3+{\cal S}_4\\
\ee 
where  ${\cal S}_i$ read as 
\begin{gather}
\label{eq:Si}
    {\cal S}_1=\frac{c}{3}  \left(\log (4)-4 T r_h e^{\frac{\delta +2 t_b}{4
   r_h}}\left(\frac{ (a-r_h)(b-r_h)
    }{ab }\right)^{\frac{1}{4}}\right),\\ \nn
    {\cal S}_2=\frac{2c}{3}  \log\! \left(\!\frac{1}{T}\,{\small\sinh 
    \frac{2 T 
    \sqrt{\sqrt{\frac{a}{r_h}-1}
   e^{\frac{a+\delta }{2 r_h}}-e^{\frac{b}{2 r_h}}
   \sqrt{\frac{b}{r_h}-1}} \sqrt{e^{\frac{a}{2 r_h}}
   \sqrt{\frac{a}{r_h}-1}-\sqrt{\frac{b}{r_h}-1} e^{\frac{b+\delta
   }{2 r_h}}}}{e^{\frac{a+b+\delta }{4 r_h}}r_h^{-3/2}\sqrt[4]{a b} }
   }\right),\\
    {\cal S}_3=\frac{c}{3}  \left(\frac{2 T r_h \sqrt{b-r_h} }{\sqrt{b}}e^{\frac{t_b}{2
   r_h}}-\log 2\right),\\
    {\cal S}_4=\frac{c}{3}  \left(\frac{2 T r_h \sqrt{a-r_h}
   }{\sqrt{a}}e^{\frac{t_a}{2 r_h}}-\log 2\right).
\end{gather}
Now we should extremize over the entanglement island parameters, i.e. the time-shift $\delta$ and the island size $a$. In the case of the ground state fields $T=0$, it was shown \cite{HIM} that the entanglement islands are located near the horizon $r_h$ and the time shift is $\delta=0$. 

\begin{figure}[t!]
\centering 
\includegraphics[width=6.5cm]{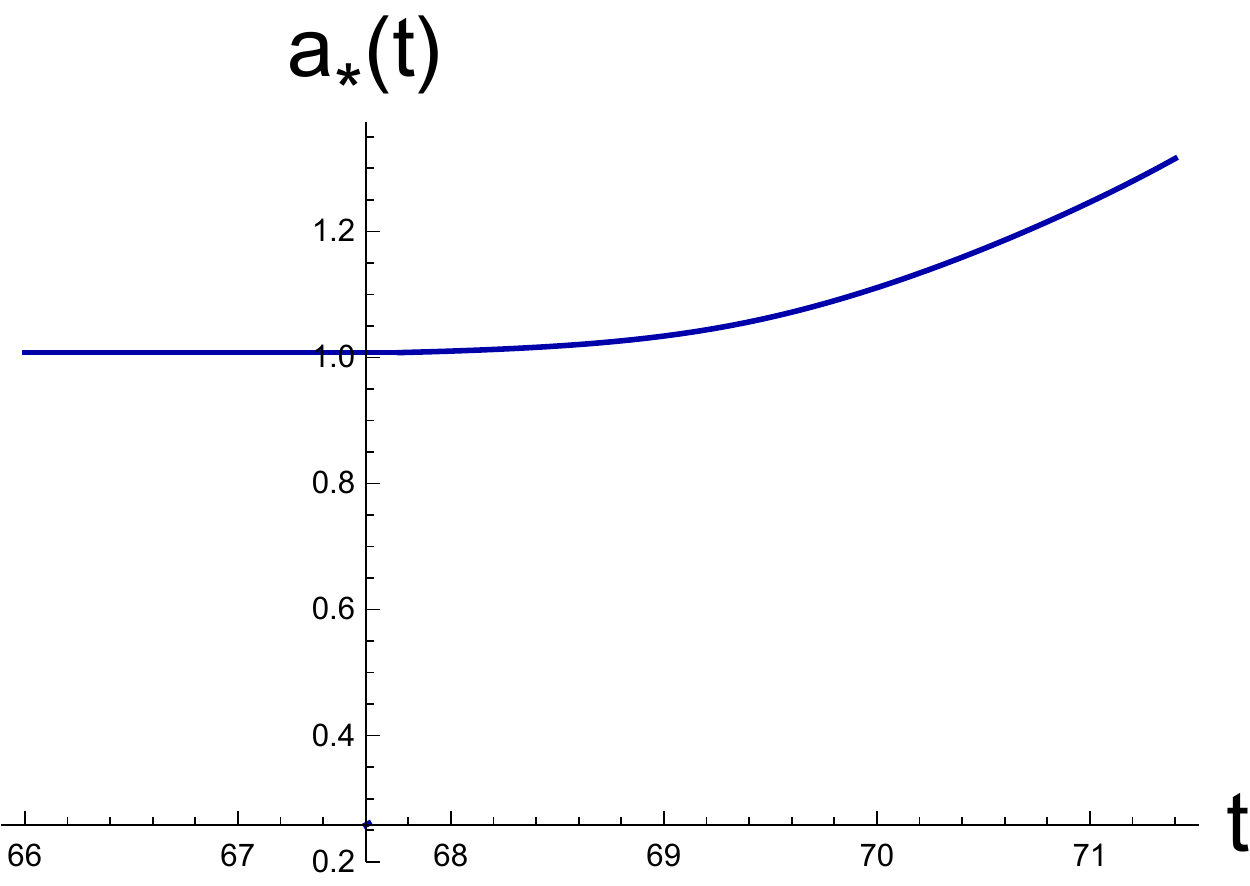}
\includegraphics[width=6.5cm]{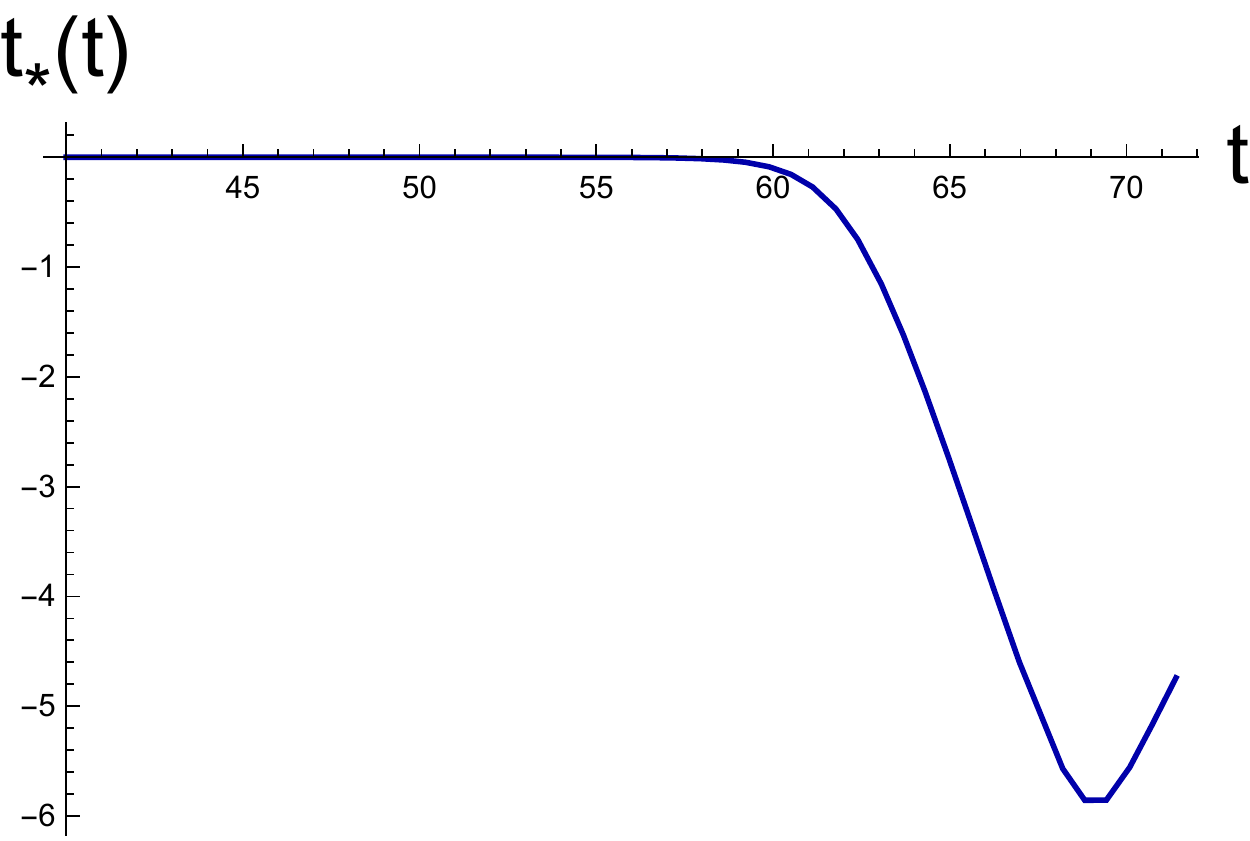}
 \caption{Time dependence of the spatial ($a_*(t)$) and time ($t_*(t)$) locations of the entanglement island. Here different necessary parameters are fixed as follows: $b=4$, $r_h=1$, $G=0.5$ and $c=3$. }
 \label{fig:AandT}
\end{figure}

Since the expressions for the entanglement entropy defined by \eqref{eq:S22} and \eqref{eq:Si} are lengthy and cumbersome, we find the extremal values $\delta_*$ and $a_*$ numerically. We present their dependence  as the function of the time $t_b$ in Fig.~\ref{fig:AandT}. One can see that at relatively early times, thermal effects are negligible, i.e. $\delta \approx 0$ and $a \approx r_h$. However, during the evolution, the spatial location of the island $a_*$ starts to move away from the horizon. The time shift $\delta_*$ at the same time shows significant deviations from the non-thermal values as well.
Substituting the extremal values $a_*$ and $t_*$ obtained from the numerical calculations, we obtain the following picture of the entanglement evolution in our setup. After the initial growth, the entanglement saturates for some time at the same value as for the ground state (i.e. when the entanglement given by \eqref{eq:Sintro} ). Being in equilibrium at this value for some amount of time, the entanglement goes to a rapid growth regime.

\subsection{Initial growth of entanglement entropy} 
The expression describing the initial growth of the entanglement entropy has a more compact form compared to the entanglement island formula. Using  \eqref{eq:S1}, after some algebra we get that the growth is described by
\be 
S_1=\frac{c}{3} \log \left(\frac{\sinh \left(\pi T\cdot \left(16 r_{h}^{2}\left(b-r_{h}\right) b^{-1}\cosh ^{2} \frac{t_{b}}{2 r_{h}}\right)^{1/2}\right)}{\pi T \varepsilon}\right).
\ee 
At early times, the growth is quadratic as it is for the zero temperature case, however, with the temperature dependent coefficients
\be 
S_1\approx \alpha+\beta\cdot t^2,
\ee 
where 
\begin{gather}
    \alpha=\log \left(\pi  \varepsilon T \cdot \text{csch}\left(\frac{4 \pi  T r_h
   \sqrt{b-r_h}}{\sqrt{b}}\right)\right),\\
    \beta=\frac{\pi  c  T \sqrt{b-r_h} \coth \left(\frac{4 \pi  T r_h
   \sqrt{b-r_h}}{\sqrt{b}}\right)}{6 \sqrt{b} r_h},
\end{gather} 
which is quite typical for the entanglement in globally perturbed states \cite{Hartman:2013qma}. After the quadratic growth as in the island phase, one can observe the intermediate regime given by a linear growth of the entanglement and coincident with the zero-temperature behavior
\be 
S_1\approx \frac{c\,t}{6\,r_h}.
\ee 
Also, as for the island phase at late times, $S_1$ exhibits a rapid exponential growth.

\subsection{Summary of the entanglement entropy behaviour of a conformal heated medium }
\begin{figure}[t]
\centering 
\includegraphics[width=7.5cm]{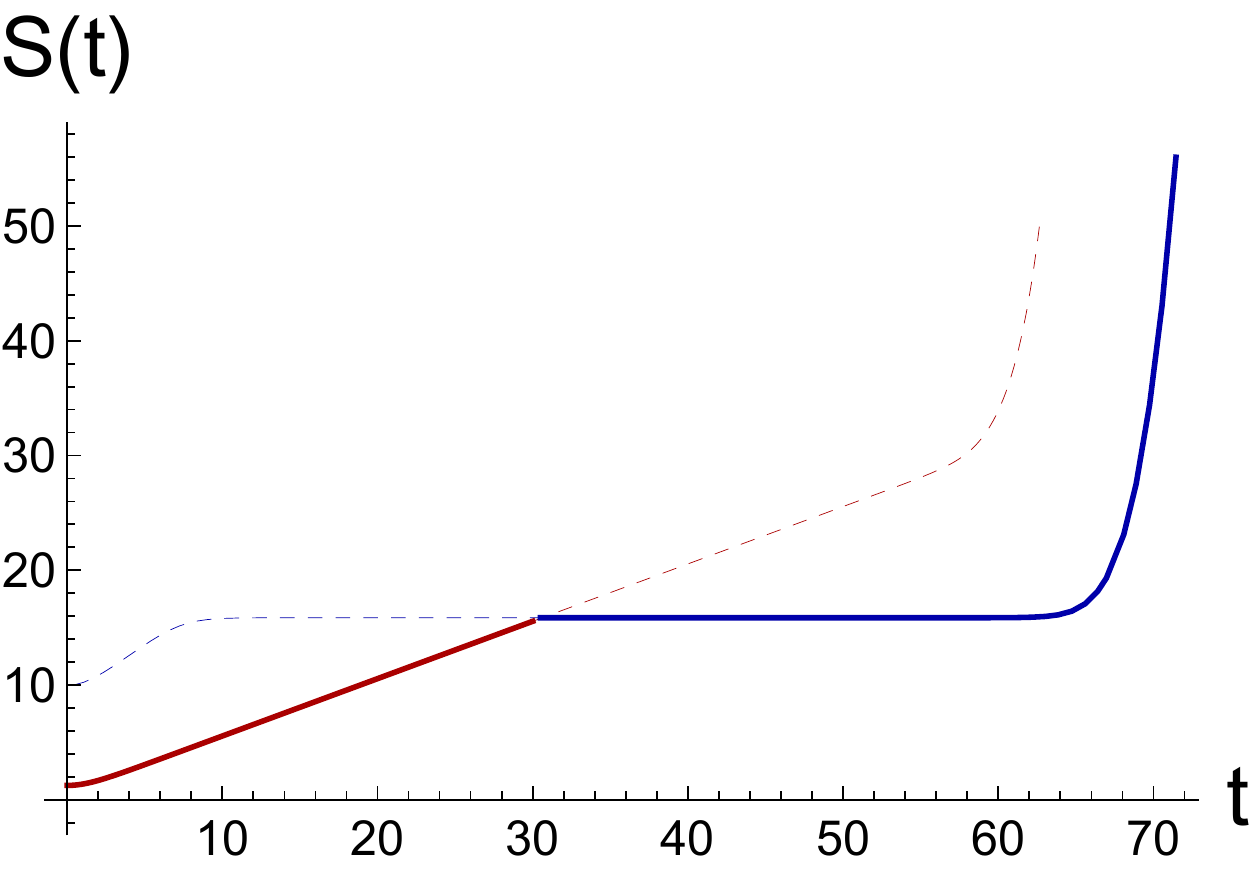}\,\,\,\,
\includegraphics[width=7.5cm]{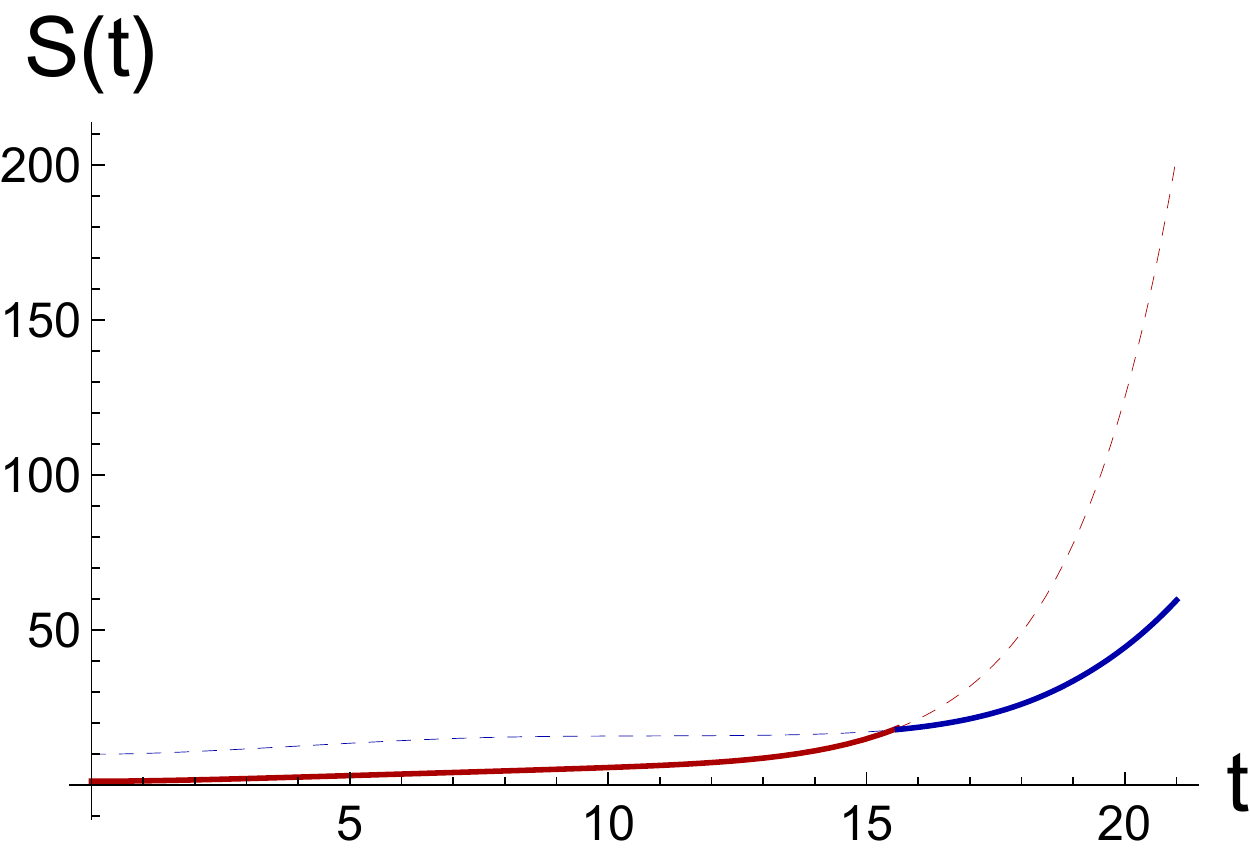}
 \caption{The entanglement entropy as the function of time $t$. The red curve (solid) corresponds to the dominant contribution given by $S_1$ and the blue one (also solid) is the contribution of the entanglement generated by the islands, while the dashed parts of the curves present the same contributions when they are not dominating. The parameters are fixed to be $G=0.5$, $r_h=1$, $c=3$, $\varepsilon=1$. The temperature for the left plot is $T=10^{-14}$, for the right one $T=0.001$. }
 \label{fig:entropy1}
\end{figure}

Finally, bringing everything together, we present a complete picture of the entanglement evolution in Fig.\ref{fig:entropy1}.
Let us briefly summarize the main features of the entanglement evolution as well as the other findings concerning the two-sided black hole surrounded by thermal fields. 
\begin{itemize}
    \item For a relatively small temperature, the entanglement entropy evolution starts with the initial quadratic growth dependent both on the black hole and matter temperature, then smoothly turns into the linear growth regime: $S \sim \gamma \cdot t $, defined solely by the black hole temperature as in the non-thermal case. After that, it saturates at some value coinciding with the one observed in the non-thermal case, and at late times, the entanglement blows up exponentially. As we mentioned before, the linear and quadratic initial growth is natural for the entanglement entropy, which can be observed in the global quench protocol in CFT described holographically by an eternal black hole or a collapsing shell.
    \item For larger CFT temperatures, the intermediate regime corresponding to a constant behavior is absent. Instead of this, the slowly growing entanglement enters the explosive growth regime directly.
    \item   An important point that makes the difference with the zero-temperature case is that the entanglement island spatial location shifts significantly for late times, approximatelyat the moment when the exponential growth of the entanglement happens. Also, it is worth noticing the presence of the non-zero time shift in the time location of the islands, i.e. $t_a-t_b \neq 0$ now.
     \end{itemize}

\section*{Concluding remarks}
\addcontentsline{toc}{section}{Concluding remarks}
\label{sec:concl}
Let us briefly summarize the obtained results,   discuss how they are related to different aspects of the information paradox/islands phenomenon and indicate future possible directions of research.
\\

We have considered a finite temperature 2d CFT in the eternal black hole and have shown how the entanglement island prescription works in such a setup. An important part of our study is that the black hole and 2d CFT have different temperatures, i.e. the CFT is considered to be described by the thermal density matrix with the temperature not equal to the black hole one. The studies of curved space density matrix and their relation to the gravitational constant renormalization have been discussed, for example, in \cite{Susskind:1994sm} (see \cite{Frolov:1998vs} for a review). It is worth noticing that the divergences in the entropy which were one of the central points of the discussion in these papers are natural in the entanglement entropy (however, they have a slightly different nature associated to the UV degrees of freedom). We find that a consideration of thermal matter spoils the behaviour of   the Page curve expected  
from  unitarity. The deviation from unitarity is reflected in the explosive growth of the entanglement at late times. For small CFT temperatures, this happens at large enough times after the period of a stationary entropy fixed by the static entanglement island. After that, the size of the island starts to grow and the generalized entropy value grows extremely fast. If we consider a non-equilibrium situation, one can naturally assume that a black hole will come to equilibrium with the CFT, therefore, the explosive period can be avoided. However, for a large enough amount of the gas, the transition to the explosive growth occurs much faster. In general, this is consistent, for example, with  the recent studies of the Hayden-Preskill protocol at finite temperature \cite{Li:2021mnl} which states that decoding of quantum information in Hawking radiation becomes much harder for the thermal density matrix under consideration.
\\

Future directions of the research include the generalization  to different quantum fields, for example, a conformal chiral matter with spin, out-of-equilibrium conformal matter (for the example given by quenched states in two-dimensional background) as well as a non-conformal matter.

\section*{Acknowledgements}
This work is supported by Russian Science Foundation
grant 20-12-00200.


\begin{thebibliography}{} 

\bibitem{Birrell:1982ix}
N.~D.~Birrell and P.~C.~W.~Davies,
``Quantum Fields in Curved Space,''

\bibitem{Haw1} S.W. Hawking,  ``Particle creation by black holes,''
Comm. Math. Phys.
43 (1975) 199

\bibitem{Haw2} S.W. Hawking, 
``Breakdown of Predictability in Gravitational Collapse,''
Phys. Rev. D \textbf{14}, 2460-2473 (1976)

\bibitem{Page:1993wv} 
  D.~N.~Page,
  ``Information in black hole radiation,''
  Phys.\ Rev.\ Lett.\  {\bf 71}, 3743 (1993)
  [hep-th/9306083]
 


\bibitem{Page:2013dx} 
  D.~N.~Page,
  ``Time Dependence of Hawking Radiation Entropy,''
  JCAP {\bf 1309}, 028 (2013)
  [arXiv:1301.4995 [hep-th]].

\bibitem{Penington:2019npb}
G.~Penington,
``Entanglement Wedge Reconstruction and the Information Paradox,''
JHEP \textbf{09}, 002 (2020)
[arXiv:1905.08255 [hep-th]].

\bibitem{Almheiri:2019psf}
A.~Almheiri, N.~Engelhardt, D.~Marolf and H.~Maxfield,
``The entropy of bulk quantum fields and the entanglement wedge of an evaporating black hole,''
JHEP \textbf{12}, 063 (2019)
[arXiv:1905.08762 [hep-th]].

\bibitem{Almheiri:2019qdq}
A.~Almheiri, T.~Hartman, J.~Maldacena, E.~Shaghoulian and A.~Tajdini,
``Replica Wormholes and the Entropy of Hawking Radiation,''
JHEP \textbf{05}, 013 (2020)
[arXiv:1911.12333 [hep-th]].



\bibitem{Penington:2019kki}
G.~Penington, S.~H.~Shenker, D.~Stanford and Z.~Yang,
``Replica wormholes and the black hole interior,''
JHEP \textbf{03}, 205 (2022)
[arXiv:1911.11977 [hep-th]].

\bibitem{Almheiri:2019yqk}
A.~Almheiri, R.~Mahajan and J.~Maldacena,
``Islands outside the horizon,''
[arXiv:1910.11077 [hep-th]].

\bibitem{Almheiri:2020cfm}
A.~Almheiri, T.~Hartman, J.~Maldacena, E.~Shaghoulian and A.~Tajdini,
``The entropy of Hawking radiation,'' Rev. Mod. Phys. \textbf{93}, no.3, 035002 (2021)
[arXiv:2006.06872 [hep-th]].

\bibitem{Chen:2020uac}
H.~Z.~Chen, R.~C.~Myers, D.~Neuenfeld, I.~A.~Reyes and J.~Sandor,
``Quantum Extremal Islands Made Easy, Part I: Entanglement on the Brane,''
JHEP \textbf{10}, 166 (2020)
doi:10.1007/JHEP10(2020)166
[arXiv:2006.04851 [hep-th]].

\bibitem{Chen:2020hmv}
H.~Z.~Chen, R.~C.~Myers, D.~Neuenfeld, I.~A.~Reyes and J.~Sandor,
``Quantum Extremal Islands Made Easy, Part II: Black Holes on the Brane,''
JHEP \textbf{12}, 025 (2020)
[arXiv:2010.00018 [hep-th]].


\bibitem{Dong:2020uxp}
X.~Dong, X.~L.~Qi, Z.~Shangnan and Z.~Yang,
``Effective entropy of quantum fields coupled with gravity,''
JHEP \textbf{10}, 052 (2020)
[arXiv:2007.02987 [hep-th]].

 \bibitem{HIM}
K.~Hashimoto, N.~Iizuka and Y.~Matsuo,
``Islands in Schwarzschild black holes,''
JHEP \textbf{06}, 085 (2020)
[arXiv:2004.05863 [hep-th]].


\bibitem{Krishnan:2020}
C.~Krishnan, V.~ Patil and  J.~Pereira, ``Page Curve and the Information Paradox in Flat Space,''
arXiv:2005.02993 [hep-th].




\bibitem{Alishahiha:2020qza}
M.~Alishahiha, A.~Faraji Astaneh and A.~Naseh,
``Island in the presence of higher derivative terms,''
JHEP \textbf{02}, 035 (2021)
[arXiv:2005.08715 [hep-th]].

\bibitem{Geng:2021wcq}
H.~Geng, Y.~Nomura and H.~Y.~Sun,
``Information paradox and its resolution in de Sitter holography,''
Phys. Rev. D \textbf{103}, no.12, 126004 (2021)
[arXiv:2103.07477 [hep-th]]


\bibitem{Matsuo:2020ypv}
Y.~Matsuo,
``Islands and stretched horizon,''
JHEP \textbf{07}, 051 (2021)
[arXiv:2011.08814 [hep-th]].

\bibitem{AV}
I.~Aref'eva and I.~Volovich,
``Quantum explosions of black holes and thermal coordinates,''
[arXiv:2104.12724 [hep-th]].



\bibitem{Wang:2021woy}
X.~Wang, R.~Li and J.~Wang,
``Islands and Page curves of Reissner-Nordstr\"om black holes,''
JHEP \textbf{04}, 103 (2021)
[arXiv:2101.06867 [hep-th]].



\bibitem{Karananas:2020fwx}
G.~K.~Karananas, A.~Kehagias and J.~Taskas,
``Islands in linear dilaton black holes,''
JHEP \textbf{03}, 253 (2021)
[arXiv:2101.00024 [hep-th]].

\bibitem{Ahn:2021chg}
B.~Ahn, S.~E.~Bak, H.~S.~Jeong, K.~Y.~Kim and Y.~W.~Sun,
``Islands in charged linear dilaton black holes,''
Phys. Rev. D \textbf{105}, no.4, 046012 (2022)
[arXiv:2107.07444 [hep-th]].

\bibitem{Ageev:2019xii}
D.~S.~Ageev, I.~Y.~Aref'eva and A.~V.~Lysukhina,
``Wormholes in Jackiw\textemdash{}Teitelboim Gravity,''
Teor. Mat. Fiz. \textbf{201}, no.3, 424-439 (2019)

\bibitem{Yu:2021cgi}
M.~H.~Yu and X.~H.~Ge,
``Page Curves and Islands in Charged Dilaton Black Holes,''
Eur. Phys. J. C \textbf{82}, no.1, 14 (2022)
[arXiv:2107.03031 [hep-th]].


\bibitem{Almheiri:2012rt}
A.~Almheiri, D.~Marolf, J.~Polchinski and J.~Sully,
``Black Holes: Complementarity or Firewalls?,''
JHEP \textbf{02}, 062 (2013)
[arXiv:1207.3123 [hep-th]].



\bibitem{Rozali:2019day}
M.~Rozali, J.~Sully, M.~Van Raamsdonk, C.~Waddell and D.~Wakeham,
``Information radiation in BCFT models of black holes,''
JHEP \textbf{05}, 004 (2020)
[arXiv:1910.12836 [hep-th]].

\bibitem{Sully:2020pza}
J.~Sully, M.~V.~Raamsdonk and D.~Wakeham,
``BCFT entanglement entropy at large central charge and the black hole interior,''
JHEP \textbf{03}, 167 (2021)
[arXiv:2004.13088 [hep-th]].

\bibitem{Akal:2020twv}
I.~Akal, Y.~Kusuki, N.~Shiba, T.~Takayanagi and Z.~Wei,
``Entanglement Entropy in a Holographic Moving Mirror and the Page Curve,''
Phys. Rev. Lett. \textbf{126}, no.6, 061604 (2021)
[arXiv:2011.12005 [hep-th]].

\bibitem{cccc}
I.~Akal, Y.~Kusuki, N.~Shiba, T.~Takayanagi and Z.~Wei,
``Holographic moving mirrors,''
Class. Quant. Grav. \textbf{38}, no.22, 224001 (2021)
[arXiv:2106.11179 [hep-th]].

\bibitem{Fernandez-Silvestre:2021ghq}
D.~Fern\'andez-Silvestre, J.~Foo and M.~R.~R.~Good,
``On the duality of Schwarzschild-de Sitter spacetime and moving mirror,'' Class. Quant. Grav. \textbf{39}, no.5, 055006 (2022)
[arXiv:2109.04147 [gr-qc]].

\bibitem{Ageev:2021iyw}
D.~S.~Ageev,
``On the entanglement contour of excited states in the holographic CFT,''
Eur. Phys. J. Plus \textbf{136}, no.4, 435 (2021)
[arXiv:1905.06920 [hep-th]].

\bibitem{Ageev:2021ipd}
D.~S.~Ageev,
``Shaping contours of entanglement islands in BCFT,''
JHEP \textbf{03}, 033 (2022)
[arXiv:2107.09083 [hep-th]].


\bibitem{mir1}
Fulling, S. A.,  Davies, P. C. (1976). Radiation from a moving mirror in two dimensional space-time: conformal anomaly. Proceedings of the Royal Society of London. A. Mathematical and Physical Sciences, 348(1654), 393-414.

\bibitem{mir2}
M.~R.~R.~Good, E.~V.~Linder and F.~Wilczek,
``Moving mirror model for quasithermal radiation fields,''
Phys. Rev. D \textbf{101}, no.2, 025012 (2020)
[arXiv:1909.01129 [gr-qc]].

\bibitem{Bianchi:2014qua}
E.~Bianchi and M.~Smerlak,
``Entanglement entropy and negative energy in two dimensions,''
Phys. Rev. D \textbf{90}, no.4, 041904 (2014)
[arXiv:1404.0602 [gr-qc]].

\bibitem{Hotta:2015huj}
M.~Hotta and A.~Sugita,
``The Fall of Black Hole Firewall: Natural Nonmaximal Entanglement for Page Curve,''
PTEP \textbf{2015}, no.12, 123B04 (2015)
[arXiv:1505.05870 [gr-qc]].

\bibitem{Good:2016atu}
M.~R.~R.~Good, K.~Yelshibekov and Y.~C.~Ong,
``On Horizonless Temperature with an Accelerating Mirror,''
JHEP \textbf{03}, 013 (2017)
[arXiv:1611.00809 [gr-qc]].

\bibitem{Chen:2017lum}
P.~Chen and D.~h.~Yeom,
``Entropy evolution of moving mirrors and the information loss problem,''
Phys. Rev. D \textbf{96}, no.2, 025016 (2017)
[arXiv:1704.08613 [hep-th]].




\bibitem{Hartman:2013qma}
T.~Hartman and J.~Maldacena,
``Time Evolution of Entanglement Entropy from Black Hole Interiors,''
JHEP \textbf{05}, 014 (2013)
[arXiv:1303.1080 [hep-th]].

\bibitem{Susskind:1994sm}
L.~Susskind and J.~Uglum,
``Black hole entropy in canonical quantum gravity and superstring theory,''
Phys. Rev. D \textbf{50}, 2700-2711 (1994)
[arXiv:hep-th/9401070 [hep-th]].

\bibitem{Frolov:1998vs}
V.~P.~Frolov and D.~V.~Fursaev,
``Thermal fields, entropy, and black holes,''
Class. Quant. Grav. \textbf{15}, 2041-2074 (1998)

[arXiv:hep-th/9802010 [hep-th]].


\bibitem{Li:2021mnl}
R.~Li and J.~Wang,
``Hayden-Preskill protocol and decoding Hawking radiation at finite temperature,''
[arXiv:2108.09144 [hep-th]].

\end{thebibliography}
\end{document}